\documentclass[10pt,conference]{IEEEtran}

\usepackage{epsfig,epsf,times,latexsym,amsfonts,amsmath,amssymb,verbatim,cite,mathrsfs}
\usepackage{graphicx}
\usepackage[dvips]{color}

\def\eE{{\mathbb E}}

\def\QED{\mbox{\rule[0pt]{1.5ex}{1.5ex}}}

\begin{document}

\title{Benefit of Delay on the Diversity-Multiplexing Tradeoffs of
MIMO Channels with Partial CSI}

\author{\authorblockN{Masoud Sharif and Prakash Ishwar}
\authorblockA{Department of Electrical and Computer Engineering \\
Boston University, Boston, MA 02215 \\ ${\rm
email:~\{sharif,pi\}@bu.edu }$}}

\maketitle

\begin{abstract}
This paper re-examines the well-known fundamental tradeoffs between
rate and reliability for the multi-antenna, block Rayleigh fading
channel in the high signal to noise ratio (SNR) regime when ($i$) the
transmitter has access to (noiseless) one bit per coherence-interval
of causal channel state information (CSI) and ($ii$) soft decoding
delays together with worst-case delay guarantees are acceptable. A key
finding of this work is that substantial improvements in reliability
can be realized with a very short expected delay and a slightly longer
(but bounded) worst-case decoding delay guarantee in communication
systems where the transmitter has access to even one bit per coherence
interval of causal CSI. While similar in spirit to the recent work on
communication systems based on automatic repeat requests (ARQ) where
decoding failure is known at the transmitter and leads to
re-transmission, here transmit side-information is purely based on
CSI. The findings reported here also lend further support to an
emerging understanding that decoding delay (related to throughput) and
codeword blocklength (related to coding complexity and delays) are
distinctly different design parameters which can be tuned to control
reliability.
\end{abstract}

\section{Introduction}\label{sec:intro}

It is well known that multiple antennas can substantially increase the
capacity of a point-to-point fading channel and can also significantly
improve the reliability of communications via space
diversity. Interestingly, both these gains can be obtained without
having any CSI at the transmitter~\cite{zhengtse,tarokh}. In fact,
even with full non-causal transmit CSI, the ergodic capacity of
multi-antenna links cannot be improved substantially at high SNR
(specifically, the scaling law of the capacity with SNR remains the
same). This motivates the question of whether any knowledge of the
channel at the transmitter could improve the reliability of
multi-antenna channels.

There is a large body of work in characterizing the reliability gains
of multiple-input multiple-output (MIMO) block-fading
channels~\cite{tarokh,arq-mimo,sandeep}. When the transmitter has no
CSI, Zheng and Tse~\cite{zhengtse} characterized the tradeoffs between
the rate and the reliability exponent at high SNRs by analyzing what
is referred to as the outage event and proving that the overall error
probability is dominated by the probability of the outage
event. Specifically, outage refers to the event that the channel
realization during the code length interval is too poor to support the
given rate~\cite{caire-delay}. It turns out that for a channel with
coherence interval $L$, the reliability exponent of a block code of
blocklength $l$ depends only on rate and the number of antennas and
does not depend on the code length $l$ (when $l$ is larger than the
total number of transmit and receive antennas). In this setting, the
decoding delay is, of course, equal to the code blocklength
$l$\footnote{Throughout the paper, we assume the code blocklength $l$
is less than or equal to the coherence interval $L$ of the
channel. Results for $l > L$ are straightforward generalizations but
are omitted. }.

Any side information which is correlated to the channel conditions
when made available to the transmitter, including causal
CSI~\cite{Shannon1958} and/or decoder status\footnote{Technically, one
distinguishes between causal CSI which is {\em independent of the
message} and is available at the transmitter {\em before} the next
transmission and {\em decoder feedback} which depends on the channel
outputs {\em after} a
transmission.}~\cite{sch-barron,burnash,yamamoto,telatar1,sahai}, can
potentially improve the rate and reliability. In particular, it is
well known that feedback can substantially improve the error exponent
of additive white Gaussian noise (AWGN) channels via variable length
coding and power control assuming no strict delays and peak power
constraints~\cite{sch-barron,burnash,yamamoto,telatar1,sahai}. For
block-fading channels, in \cite{arq-mimo} an automatic retransmission
request (ARQ) is shown to substantially improve the diversity gain by
allowing codeword retransmissions (random variable-length channel
codes) with the aid of noiseless one-bit decision feedback and power
control. Therefore, in a communication system with feedback, the
decoding time $T$ ($\#$channel uses) is a random variable depending on
$l$ and the state of the decoder. In order to capture this random
decoding delay, Burnashev introduced the notion of error exponents for
a feedback code which transmits one of $M$ messages in {\em expected
time} at most $l$~\cite{burnash}.

In this paper, our focus is on the former type of transmitter side
information, namely causal CSI. In particular, we assume that the
transmitter has access to only one noiseless bit of causal side
information per coherence interval from the receiver which describes
the channel state as being ``good'' (above a predecided threshold) or
bad (below the threshold). This type of channel state feedback has
substantially less coding complexity than that of the ARQ systems as
it only requires the knowledge of the channel state and not the status
of the decoder. Following~\cite{burnash}, we define the reliability
function as the smallest error probability that can be achieved by a
code of length $l$ which transmits one of $\rho^{rl}$ messages in the
expected time at most $(l+\epsilon)$ where $r$ is the multiplexing
gain and $\epsilon$ tends to zero as the SNR $\rho$ goes to
infinity. By leveraging this side information and the notion of an
expected decoding delay, we show that the reliability function can be
substantially improved in the high SNR regime.  In our scheme,
transmission occurs only when the channel satisfies a certain
condition which can be ascertained at the receiver. Therefore, similar
to feedback channels, the decoding delay is random and rate may be
reduced as we do not use the channel all the
time~\cite{sahai,arq-mimo,yamamoto}. However, it turns out that in the
regime of high SNR, the channel most likely satisfies the conditions
and thus the scaling law of the rate remains unchanged.

When there is no delay constraint and noiseless one-bit causal CSI
feedback is available, we show that the outage events can be
completely avoided and an error exponent which is as good as that of a
pure multi-antenna AWGN channel without fading can be attained. In
this case, in sharp contrast to the case where the transmitter has no
CSI, the error probability approaches zero exponentially with the {\em
blocklength} for all multiplexing gains. We further consider the case
where the decoding delay has a strict deterministic deadline, that is,
a finite maximum delay constraint, in which the receiver has to decode
the message in at most $D$ coherence intervals. In this case, it is
clear that outage events cannot be avoided and the effects of fading
will reappear. However, we show that a substantial gain on the
reliability function can be achieved even with a specified finite
maximum decoding delay. For instance, in a channel with $M$ transmit
antennas, and one receive antenna, $\%90$ of the diversity gain of an
AWGN channel can be attained with a worst case delay of
$D=\log_M{(10l/M)}$. This shows when $l=10$ and $M=3$, we only need a
delay of $D=4$ to completely overcome the effects of fading and
approach the performance of MIMO AWGN channels.

This paper is organized as follows: In Section~\ref{sec:bkgnd}, we
present the channel model and introduce the problem set-up. In
Section~\ref{sec:inftydelay}, we obtain the reliability exponent when
there is no bound on the worst case delay. In
Section~\ref{sec:findelay}, we analyze the case where the maximum
decoding delay is bounded and treat the scenarios with and without a
finite peak to average power constraint.

\section{Background}\label{sec:bkgnd}

\subsection{Channel Model}\label{sec:chmdl}

We consider a frequency-flat block-fading MIMO channel with $M$
transmit and $N$ receive antennas. The channel is assumed to remain
fixed for $L$ channel uses (the channel coherence interval) and change
independently to another state in the next block. Therefore, within
each coherence interval, the received signal at time $t$ can be
expressed as,
\begin{equation}
y_t= \sqrt{\frac{\rho}{M}} H S_t+W_t, ~~~~~~~t=1,\ldots,L,
\end{equation}
where $\rho$ is the average transmit power, $H$ denotes the (fixed)
$N\times M$ random fading channel matrix, and for each discrete time
$t$, $S_t$ denotes the $M\times 1$ channel input symbol and $W_t$ the
$N\times 1$ additive white Gaussian noise vector. All the entries of
$H$ and $W_t$ are assumed to have a zero-mean, unit-variance, complex
circular Gaussian distribution. Furthermore, the transmit message
$S_t$ at each time $t$ is required to satisfy the average power
constraint $\eE(S^\dag_tS_t)\leq 1$. In all random coding bounds for
the reliability function, we assume $S_t$ to be a random Gaussian
codeword with blocklength $l \leq L$. The $l > L$ case can also be
analyzed but is omitted for brevity.

\subsection{Reliability, Rate: No CSI at the Transmitter}

Rate versus reliability tradeoffs have been well studied for
point-to-point AWGN channels through the notions of the error exponent
(large blocklength) and the diversity gain (large SNR). For fading
MIMO channels with no transmit CSI, Zheng and Tse studied the smallest
error probability that can be achieved in the asymptotic of high SNR
by a code of size $\rho^{rl}$. In this regime, the error probability
is characterized by the diversity gain defined as,
\begin{equation}
  d(r)=\underset{\rho \rightarrow \infty}{\lim\inf} \frac{-\log{\eE_{H}\left\{P_{e}(R,\rho,l,M,N)\right\}}}{\log{\rho}}
\end{equation}
where $R=\frac{\log{\rho^{rl}}}{l}$ is the code rate in bits per
channel use, $l$ is the code blocklength, and $P_{e}$ is the smallest
achievable error probability for a rate-$R$, blocklength $l$ code. It
is further shown that when the transmitter has no CSI, $d(r)$ is given
by a piecewise linear function connecting the points $(r,(M-r)(N-r))$
for $r=0,1,\ldots,\min(M,N)$. More explicitly,
\begin{equation}
d(r)=\alpha_{k}-r\beta_{k}, ~~~~~~k-1\leq r \leq k, \label{eq:nnooo}
\end{equation}
where $1\leq k \leq \min(M,N)$, $\alpha_{k}=MN-2k(M+N)+3k^2-k,$ and
$\beta_{k}=M+N-2k+1$.

It is interesting to note that unlike AWGN channels, the error
probability of fading channels (without transmit CSI) tends to zero
with an exponent {\em independent of the code blocklength} as long as
$l \geq (M+N)$. This is due to the fact that the error probability is
dominated by the so-called outage error event: the event of having a
sequence of atypically poor channel fading gains over the duration of
a codeword. In this setting, in~\cite{zhengtse} it is shown that the
optimal diversity gain can be achieved by using a fixed-length
blockcode and when the {\em decoding delay is equal to the length of
the codeword blocklength} $l$.

\subsection{Reliability, Rate, and Decoding Delay: Transmit CSI}

Transmit side information in its broadest sense includes both CSI and
the received signal at the decoder (feedback channels). It is folklore
that while feedback cannot improve the capacity of a discrete
memoryless channel, it can improve the reliability function (see
\cite{sch-barron,Shannon1958} and references therein). The gain in the
reliability is achieved by leveraging ($i$) the possibility of
re-transmissions while keeping the codeword blocklength fixed, and
($ii$) using power control to boost the power when rare events happen
\cite{sch-barron}. This implies that the decoding time $T$ is random
and no longer equals the codeword length. For general DMCs,
in~\cite{burnash}, Burnashev obtained tight bounds on the smallest
error probability of a code which transmits one of the messages in
expected time at most $l$, that is, $\eE(T)\leq l$ and unbounded
maximum delay~\cite{yamamoto,burnash}. In the asymptotic of large SNR,
Gamal, Caire, and Damen proposed an ARQ scheme and proved that a
single noiseless bit of feedback per codeword block pertaining to the
status of the decoder can substantially improve the reliability even
when a finite maximum decoding delay constraint is imposed, that is,
with a finite total number of retransmissions.

Inspired by these results on feedback channels, we explore the
high-SNR asymptotics of the channel reliability function with one
noiseless bit of causal CSI at the transmitter. We show that with this
limited transmit CSI and for a fixed codeword blocklength, the
reliability function can be substantially improved even with a strict
finite maximum decoding delay constraint.

Here, the reliability function is defined in terms of the smallest
achievable error probability for a blockcode of blocklength $l$, size
$\rho^{rl}$, and maximum decoding time of $lD$ (that is, $T\leq
lD$). In particular, following~\cite{burnash}, we define the diversity
gain with CSI as
\begin{equation}
  d_{1-bit}({r},D)=\underset{\rho \rightarrow \infty}{\lim\inf}
  \frac{-\log{\eE_{H}\left\{P_{e}({R},\rho,l,
  M,N,D)\right\}}}{\log{\rho}},
\label{eq:def-rel}
\end{equation}
where ${R}=\frac{\log{\rho^{rl}}}{\eE(T)}$, $T$ is decoding time, and
the multiplexing gain $r$ is defined as,
\begin{equation}
{r}=\underset{\rho \rightarrow \infty}{\lim}\frac{R}{\log{\rho}}.
\label{defr}
\end{equation}

In this paper we assume that the transmitter knows, through one bit
receiver feedback, whether or not the channel realization satisfies a
predecided criterion before transmission. The transmitter leverages
this side information to postpone the transmission upto the time that
the channel is favorable. When $D$ is bounded and the transmitter has
delayed the transmission for $(D-1)$ coherence blocks, the message
will be sent in the next coherence block. In what follows, we
investigate the behavior of $d_{1-bit}(r,D)$ for different values of
$D$.

\section{Diversity-Multiplexing Tradeoff: Bounded Expected Delay,
  Unbounded Maximum Delay} \label{sec:inftydelay}

The reliability of MIMO fading channels without transmit CSI is
dominated by the probability of the outage
event~\cite{zhengtse}. However, in the asymptotic of large SNR, the
outage event is a rare event with a small probability of
$\rho^{-d(r)}$ where $d(r)$ is defined in (\ref{eq:nnooo}).  Therefore
knowing whether or not the channel is in outage, messages can be
scheduled for transmission only during favorable conditions. This
would eliminate the atypical outage events and may also lead to
unbounded decoding delay, though the expected delay can still be
bounded. This is due to the fact there is a nonzero probability,
however small, of arbitrarily long sequence of outage events.

When there is no maximum decoding delay constraint, we assume that the
receiver sends one bit to the transmitter at the beginning of each
coherence interval indicating the occurrence/non-occurrence of the
event:
\begin{equation}
  {\cal O}_{r,\infty}:   \log\det (I+\rho H^*H) \leq
  \min(M,N)(\log{\rho}-2\log\log{\rho}).\nonumber
\end{equation}
A message is transmitted only when the channel fails to satisfy the
condition ${\cal O}_{r,\infty}$. The following theorem provides the
reliability function defined in (\ref{eq:def-rel}) with $D$ infinite.

{\bf Theorem~1}~{\em (Lower bound on diversity gain) Consider the
channel of Section~\ref{sec:chmdl} with noiseless one-bit causal
transmit CSI confirming or denying condition ${\cal O}_{r,\infty}$.
For any blocklength-$l$, $l\leq L$, there exists a block code with
$\rho^{rl}$ codewords for which
\begin{equation}
 d_{1-bit}(r,\infty) \geq l (\min(M,N)-r),
\label{eq:nodelay}
\end{equation}
where $d_{1-bit}$ is as defined in (\ref{eq:def-rel}).}

{\bf Remark:} It is interesting to note that the error probability is
now decreasing exponentially with the blocklength unlike the case
where there is no transmit CSI. Comparing this error exponent with
that of an AWGN channel with the same SNR, it is seen that the effect
of channel fades can be completely removed. This substantial gain is
obtained due to the relaxed (but practical) requirement on the
decoding delay from being exactly equal to the codeword blocklength to
being equal to the expected number of channel uses until codeword
reception. It is straightforward to establish that the average
decoding delay is equal to
$l(1+\Theta(\frac{1}{(\log{\rho})^2}))\approx l$.

{\em Proof-sketch:} By bounding the determinant of a matrix by its
minimum eigenvalue and its trace, it readily follows that ${\rm
Pr}({\cal
O}_{r,\infty})=\Theta\left(\frac{1}{(\log{\rho})^2}\right)$. This
implies that the decoding delay $T$ has a geometric distribution with
parameter $1-{\rm Pr}({\cal O}_{r,\infty})$. Therefore, the
multiplexing gain as defined in (\ref{defr}) is equal to,
\begin{equation}
    \underset{\rho \rightarrow \infty}{\lim} \frac{R}{\log{\rho}}=\underset{\rho \rightarrow \infty}{\lim}  \frac{rl\log{\rho}}{l \log{\rho}/(1-{\rm Pr}({\cal O}_{r,\infty}))}=r.
  \end{equation}
An achievable error probability can be found using the random coding
bound. Following~\cite{tarokh}, we may write,
  \begin{eqnarray}
  \eE_{H}\{P_{e}\}&\leq & \hspace{-2mm} \rho^{rl} \int_{{\cal O}_{r,\infty}^c} \frac{f_{H}(H)dH}{( \det(I+\rho H^*H))^{l}}  \nonumber \\
     & \leq & \rho^{-l (\min(M,N)-r)} (\log{\rho})^{2l}, \nonumber
  \end{eqnarray}
where ${\cal O}_{r,\infty}^{c}$ is the complement of the event ${\cal
O}_{r,\infty}$. \QED

In the following section, we investigate the scenario in which the
message decoding delay is bounded.

\section{Diversity-Multiplexing-Delay Tradeoff: Bounded Maximum Delay}
\label{sec:findelay}

In the previous section, we observed that a substantial gain can be
obtained in the diversity gain with little CSI at the transmitter at
the cost of having the possibility of infinite decoding delays (albeit
with arbitrarily small probability for large SNR). In this section, we
further impose a worst case delay constraint of $D$ for message
decoding. In other words, we assume that the decoder has to decide on
a message at most in $D$ coherence intervals of the channel. Clearly,
this would imply that there is a non-zero probability for a sequence
of really bad channel events.

When $D=1$, the problem reduces to the problem considered
in~\cite{zhengtse} with the only difference that the transmitter has
1-bit causal CSI. Since there is no possibility for postponing the
transmission and power allocation optimization for one coherence
interval does not change the diversity gain, it is straightforward to
show that $d_{1-bit}(r,1)=d(r)$\footnote{The reason being that in the
regime of large SNR, the exponent of the scaling law of the outage
probability with SNR would not be changed by optimal power allocation
over one coherence interval.} .

In this section, we prove that with only one bit causal CSI feedback
at the transmitter, one can substantially improve the diversity gain
of codes of length $l$ by leveraging the possibility of postponing the
transmission and/or exploiting the long-term average power
constraint. This substantial gain is achieved by only transmitting
when the channel is in a favorable condition or we have reached the
maximum transmission delay. Since the event of having a sequence of
unfavorable channels is very unlikely, one can, without violating a
long-term average (over messages) power constraint, boost the transmit
power when faced with greater delays. We also explore the potential
gains for a (short-term) peak to average power constraint.

\subsection{Short-Term Average Power Constraint}

Here we assume that the receiver sends one bit feedback to the
transmitter confirming or denying the following condition:
\begin{equation}
{\cal O}_{r,D}:  \log\det (I+\rho H^*H) \leq f(r,D) \log{\rho}.
\label{eq:odef}
\end{equation}
where $f(r,D)$ is defined as,
\begin{equation}
f(r,D)= r + \frac{d(r) + (D-1)M_{eq}(\min(M,N)-r)}{l + (D-1)M_{eq}},
\label{eq:def-f}
\end{equation}
where $l$ is the code blocklength, $M_{eq}=|M-N|+1$, and $d(r)$ is as
defined in (\ref{eq:nnooo}). Of course, this is the outage event and
has a probability of $\rho^{-d(f(r,D))}$. In our transmission scheme,
the transmitter postpones the transmission until the channel belongs
to the set of channels that do not satisfy the condition ${\cal
O}_{r,D}$ or reaches the maximum delay constraint of $D$. In this
transmission scheme, transmission always occurs with a transmit power
$\rho$.

{ \bf Theorem~2}~{\em For the channel of Section~\ref{sec:chmdl}, the
diversity gain of a code of blocklength $l$ and cardinality
$\rho^{rl}$, with short-term power constraint $\rho$, is at least
equal to
\begin{eqnarray*}
&& d_{1-bit}(r,D) ~~ \geq ~~ l(f(r,D) - r) ~~ = ~~ \\
&& \frac{l}{l + (D-1)M_{eq}}(d(r) + (D-1)M_{eq}(\min(M,N)-r))
\end{eqnarray*}
for any $(M+N) \leq l \leq L$, where $M_{eq}=|M-N|+1$, $0 \leq r \leq
\min(M,N)$, and $d(r)$ is as defined in (\ref{eq:nnooo}).}

{\em Proof-sketch:} We use the random coding bound to obtain an upper
bound on the error probability. We condition the error probability
into two events, namely, $i)$ ${\cal A}_1$: the channel realization
satisfies ${\cal O}_{r,D}^c$ in at least one out of the first $(D-1)$
coherence intervals, $ii)$ ${\cal A}_1^c$: the channel satisfies
${\cal O}_{r,D}$ during all $(D-1)$ coherence intervals. The error
probability in the event of ${\cal A}_1$ can be easily bounded using
the same approach as in Theorem~1 as,
\begin{equation}
\eE(P_{e,{\cal A}_1}) \leq \rho^{l(r-f(r,D))}.
\end{equation}
Also, note that the probability of the event ${\cal A}_1^{c}$ is the
event that condition ${\cal O}_{r,D}$ (cf.~(\ref{eq:odef})) is met
$(D-1)$ times. Therefore,
\[ {\rm Pr}({\cal A}_1^c) \leq \rho^{-(D-1)d(f(r,D))}. \]
Furthermore, conditioned on the event ${\cal A}_{1}^c$, the error
probability is given by the error probability for $D=1$.  Therefore,
\begin{eqnarray}
\eE(P_{e}) & = & \eE(P_{e,{\cal A}_1})+ \eE(P_{e,{\cal A}_1^c})  \nonumber \\
   & \leq  & \rho ^{l(r-f(r,D))} + \rho^{-(D-1)d(f(r,D))-d(r)}.
   \label{eq:bound-mas}
\end{eqnarray}
In order to calculate the largest achievable diversity gain, one needs
to minimize the upper bound on the error probability in
(\ref{eq:bound-mas}), that is, match the exponents by finding $f(r,D)$
which satisfies,
\begin{equation}
l(r-f(r,D))=-(D-1)d(f(r,D))-d(r) \label{eq:fsol}.
\end{equation}
The expression in (\ref{eq:def-f}) solves (\ref{eq:fsol}) with
$d(f(r,D))$ replaced by the smaller value
$M_{eq}(\min(M,N)-f(r,D))$. This corresponds to minimizing a larger
upper bound (than in (\ref{eq:bound-mas})) on the error
probability. Replacing $f(r,D)$ in (\ref{eq:bound-mas}) completes the
proof. \QED

\begin{figure}
\begin{center}
\includegraphics[scale=0.5]{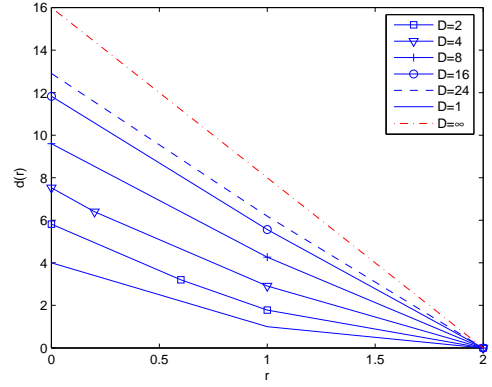}
\end{center}
\caption{\small \sl Diversity-multiplexing tradeoffs with one-bit
  noiseless causal transmit CSI for different values of $D$ for
  $M=N=2$ and $l=8$.} \label{fig:fig2}
\end{figure}

Specializing Theorem~2 to the case $N=1$ gives,
\begin{equation}
d_{1-bit}(r,D) \geq \frac{l(MD-1)}{l+M(D-1)}(1-r),
\label{eq:exmp2}
\end{equation}
for any $0\leq r \leq 1$. This shows that when $MD \gg l$, $d_{1-bit}
\gtrsim l(1-r)$ and when $l \gg MD$, $d_{1-bit} \gtrsim (MD-1)(1-r)$.
Figure~\ref{fig:fig2} shows the diversity multiplexing tradeoff for
increasing values of $D$. It is clear that with little side
information and a small delay tolerance, the reliability function can
be significantly improved. In the next section, we show that the gains
with bounded delay can be further increased by exploiting a long-term
power constraint, that is, by boosting the power when rare events
happen.

\subsection{Long-Term Average Power Constraint}

In the regime of high SNR, the outage event has a very small
probability and therefore the transmitter can boost the power in
proportion to the corresponding probability when the transmitter is
subject to a long-term power constraint. In this section, we show that
we can further improve the reliability function using power control.

Side information here is assumed to be one bit causal CSI that informs
the transmitter whether the following condition is true or false:
\begin{equation}
  {\cal O}_{r,D,i}:  ~~~~~ \log\det (I+\rho^{g_i(r,D)} H^*H) \leq f_i(r,D) \log{\rho}.
\end{equation}
where $i$ denotes the number of consecutive channels that satisfies
the conditions $ {\cal O}_{r,D,1}, \ldots, {\cal O}_{r,D,i-1}$,
respectively. Here $f_i(r,D)$ and $g_i(r,D)$ are defined as,

\begin{eqnarray}
f_i(r,D) \hspace{-2mm} & = & \hspace{-2mm}
\frac{(l-\beta_{k})r-\frac{M}{M-1}\left(\sum_{i=1}^{D}M_{eq}^{i}
-D\right) -\alpha_{k}}{l+\frac{M}{M-1}\left(\sum_{i=1}^{D}M_{eq}^{i}
-D\right)(1+\alpha_{k})}, \nonumber \\
g_i(r,D) \hspace{-2mm} & = & \hspace{-2mm} 1+\sum_{j=1}^{i-1}{\rm
Pr}(O_{r,D,j}).
\end{eqnarray}
where $g_1(r,D)=1$. Here $g_i(r,D)$ is the transmit power when the
previous $i-1$ channels satisfy all the conditions of ${\cal
O}_{r,D,1}, \ldots,{\cal O}_{r,D,i-1}$. The next theorem provides an
achievable upper bound on the error probability that implies a lower
bound for the diversity gain.

{\bf Theorem~3}~{\em For the channel of Section~\ref{sec:chmdl}, the
diversity gain of a code of blocklength $l$ and cardinality
$\rho^{rl}$, with long-term average power constraint $\rho$, is at
least equal to
\begin{eqnarray}
   d_{1-bit}(r,D)\geq \frac{l}{l+D_{eq}M_{eq}}\left(\min(M,N)
  M_{eq}D_{eq} \right. \nonumber \\
  \left. +\alpha_{k}-(D_{eq}+\beta_{k})r\right), \nonumber
\end{eqnarray}
where for $M\neq N$,
\begin{equation}
D_{eq}= \frac{1}{M-1}\left( \sum_{i=1}^{D}M_{eq}^{i}-D \right)-1,
\end{equation}
and for $M=N$, we have $D_{eq}=\frac{D(D+1)}{2}$.}

{\em Proof-sketch:} The proof follows again using the random coding
upper bound on the error probability.  The only advantage here is that
the probability of not transmitting until the $D$'th coherence
interval is much smaller as we can use power control. This can
significantly improve the achievable diversity gain. \QED

In the special case where $N=1$, Theorem~3 implies that the achievable
diversity gain is no smaller than
\begin{equation}
  d_{1-bit}(r,D)\geq \frac{l \frac{M}{M-1}\left(\sum_{i=1}^{D} M^{i} -D\right)}{l +\frac{M}{M-1}\left(\sum_{i=1}^{D} M^{i} -D\right)-M}  (1-r).
\label{eq:exmp1}
\end{equation}
Figure~\ref{fig:fig2} compares the achievable diversity gains with and
without power control as obtained in (\ref{eq:exmp1}) and
(\ref{eq:exmp2}), respectively. It is clear that with only $D=4$, we
can obtain most of the gains achieved via infinite worst case delay
obtained in (\ref{eq:nodelay}). We can quantify this observation by
computing the delay $D$ required to achieve within $\epsilon$ of the
lower bound for $d_{1-bit}(r,\infty)$, that is,
$l(1-\epsilon)(1-r)$. It is straightforward to show that the delay
required is equal to
\begin{enumerate}
\item $D=\frac{l}{\epsilon M}$ with no power control,
\item $D=\log_{\scriptscriptstyle M}\!\!\left({\frac{l}{\epsilon
M}}\right)$ with power control.
\end{enumerate}
Therefore with power control, we need an exponentially smaller delay
to achieve the same performance as opposed to the case where no power
control is employed.

\begin{figure}
\begin{center}
\includegraphics[scale=0.5]{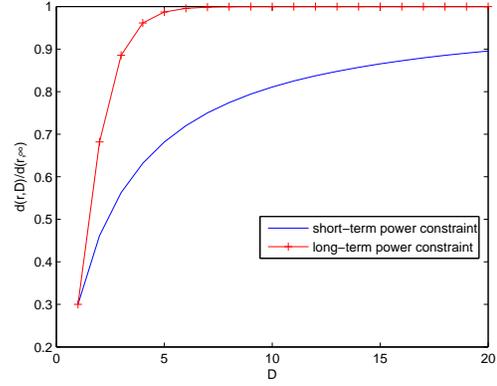}
\end{center}
\caption{\small \sl For a system with $M=3$, $N=1$, and $l=10$, the
figure shows $\frac{d_{1-bit}(r,D)}{d_{1-bit}(r,\infty)}$ with respect
to $D$ for any $r$ and for the two cases with long-term/short-term
power constraints.} \label{fig:fig3}
\end{figure}

\section*{Acknowledgment}
The authors would like to thank Prof.~S.S.~Pradhan, EECS UMich Ann
Arbor, for fruitful discussions and comments.

\end{document}